\documentclass[12pt]{article}

\usepackage{cite}
\usepackage{epsfig}
\usepackage{amsmath}
\usepackage{pstricks}

\def\mathswitch#1{\relax\ifmmode#1\else$#1$\fi}
\def\mathswitchr#1{\relax\ifmmode{\mathrm{#1}}\else$\mathrm{#1}$\fi}

\newcommand{\gev}{\,\, \mathrm{GeV}}

\newcommand{\RR}{{\rm R}}

\newcommand{\anc}{\rule{0mm}{0mm}}

\newcommand{\gesim}{\,{_{\textstyle
>}\atop^{\textstyle\sim}}\,}

\newcommand{\neu}{\tilde{\chi}^0}
\newcommand{\cha}{\tilde{\chi}}
\newcommand{\sm}{\tilde{\mu}_\RR}
\newcommand{\se}{\tilde{e}_\RR}
\newcommand{\sll}{\tilde{\ell}_\RR}

\newcommand{\mycaption}[1]{\caption{\sl #1}}
\makeatletter
\renewcommand\section{\@startsection {section}{1}{\z@}%
                                   {-3.5ex \@plus -1ex \@minus -.2ex}%
                                   {2.3ex \@plus.2ex}%
                                   {\normalfont\large\bfseries}}
\renewcommand\subsection{\@startsection{subsection}{2}{\z@}%
                                     {-3.25ex\@plus -1ex \@minus -.2ex}%
                                     {1.5ex \@plus .2ex}%
                                     {\normalfont\normalsize\bfseries}}
\makeatother

\begin{document}
\thispagestyle{empty}

\def\thefootnote{\fnsymbol{footnote}}

\begin{flushright}
\end{flushright}

\vspace{1cm}

\begin{center}

{\Large\sc {\bf Feasibility of slepton precision measurements\\
 at a muon collider}}
\\[3.5em]
{\large\sc
A.~Freitas
}

\vspace*{1cm}

{\sl 
Department of Physics \& Astronomy, University of Pittsburgh,\\
3941 O'Hara St, Pittsburgh, PA 15260, USA
}

\end{center}

\vspace*{2.5cm}

\begin{abstract}
Detectors at a high-energy muon collider must be protected from  the decay
products of beam muons by installing shielding material around
the beam pipe.
In this article, the impact of these blind detector regions
on new-physics signatures with invisible final-state particles is shown
by studying the production of sleptons, the superpartners of leptons.
Special attention is given to large backgrounds from two-photon collisions.
It is demonstrated how the influence of these backgrounds can be controlled
by implementing suitable cuts or running at two different
center-of-mass energies, thus permitting precision measurements 
of the mass and spin of the sleptons.
However, these methods become ineffective
for small mass differences between the sleptons and their decay
products, and it will be difficult to analyze the slepton signal at a muon collider
in this case.
\end{abstract}

\setcounter{page}{0}
\setcounter{footnote}{0}

\newpage


\section{Introduction}

A high-energy muon collider (MC) is envisaged to be an ideal environment for
performing precision measurements in the multi-TeV regime. One major advantage
is its sharply defined collision energy, which is only mildly distorted by
initial-state radiation, while the influence of beamstrahlung can be completely
neglected for most purposes \cite{mc}. However, the design of the detector and
beam-delivery system will have to account for the large flux of electrons from muon
decay inside the beam pipe. The detector can be protected from these decay
electrons by installing cone-shaped tungsten shields around the beam pipe near 
the interaction point.
Nevertheless, low-energy secondaries produced in the shielding material may still 
penetrate the detector, leading to large occupancy values in the inner
detector elements. Simulations show that this background can be reduced
to acceptable levels 
by using high-resolution timing information (since secondary
particle showers will arrive in the detector with some time delay) and
sufficiently massive shielding around the beam pipe\footnote{The reader is referred to the
literature for more details \cite{beamback}.}.

However, the absence of instrumentation inside the shielding cones leads to
reduced angular coverage of the
detector, which can affect the effectiveness of new physics searches and
measurements. 
In particular, physics signatures with missing energy can receive large
backgrounds from photon-photon interactions\footnote{Although it was recognized
earlier that the two-photon background is important, see $e.\,g.$
Ref.~\cite{klasen}, it has not been considered in pertinent previous studies on this
topic \cite{mcued}.}. Here, the photons are radiated off
the incident beam muons, which subsequently escape into the blind cone regions,
leading to apparent missing energy and momentum, see Fig.~\ref{schem}.

\begin{figure}
\vspace{3cm}
\centering\anc
\pspolygon*[linecolor=lightgray](4,2)(4,0.3)(0.6,0.3)
\pspolygon*[linecolor=lightgray](4,-2)(4,-0.3)(0.6,-0.3)
\pspolygon*[linecolor=lightgray](-4,2)(-4,0.3)(-0.6,0.3)
\pspolygon*[linecolor=lightgray](-4,-2)(-4,-0.3)(-0.6,-0.3)
\psline[linewidth=0.5pt](4,0.3)(4,3)(-4,3)(-4,0.3)
\psline[linewidth=0.5pt](4,-0.3)(4,-3)(-4,-3)(-4,-0.3)
\psline[linewidth=1pt, arrowsize=4pt 3]{->}(-6,0)(-1,0)(5,1)
\psline[linewidth=1pt, arrowsize=4pt 3]{->}(6,0)(1,0)(-5,1)
\pscurve[linewidth=1pt, arrowsize=4pt 2]{->}(-1,0)(-0.9,-0.2)(-0.75,-0.125)(-0.6,-0.05)
(-0.5,-0.25)(-0.4,-0.45)(-0.25,-0.375)(-0.15,-0.425)(0,-0.5)
\pscurve[linewidth=1pt, arrowsize=4pt 2]{->}(1,0)(0.9,-0.2)(0.75,-0.125)(0.6,-0.05)
(0.5,-0.25)(0.4,-0.45)(0.25,-0.375)(0.15,-0.425)(0,-0.5)
\psline[linewidth=0.5pt]{->}(0,-0.6)(0.1,-2.5)
\psline[linewidth=0.5pt]{->}(-0.05,-0.6)(-0.5,-2)
\psline[linewidth=0.5pt]{->}(0.05,-0.6)(0.7,-2)
\psline[linewidth=0.5pt]{->}(0.03,-0.37)(0.2,2)
\psline[linewidth=0.5pt]{->}(-0.03,-0.37)(-0.3,2.2)
\rput(-5.5,1){$\mu^+$}
\rput(5.5,1){$\mu^-$}
\rput(-0.6,-0.7){$\gamma$}
\rput(0.6,-0.7){$\gamma$}
\vspace{3cm}
\mycaption{Schematic depiction of conical shielding cones in a MC
detector, and a two-photon event with fake missing energy and
momentum.}
\label{schem}
\end{figure}

In this paper, the effect of the blind cones on the precision
analysis of sleptons, the supersymmetric partners of leptons, is investigated.
For concreteness, production of R-smuons $\sm$ and R-electrons $\se$ is
considered, and it is assumed that they decay directly to the lightest
neutralino $\neu_1$, which is assumed to be the lightest supersymmetric particle and thus
stable and invisible to the detector:
\begin{align}
\mu^+\mu^- &\to \sm^+\sm^- \to \mu^+\mu^-\neu_1\neu_1, \label{smu} \\
\mu^+\mu^- &\to \se^+\se^- \to e^+e^-\neu_1\neu_1. \label{sel}
\end{align}
It is demonstrated that Standard Model (SM) backgrounds, especially from
two-photon interactions, significantly affect the possibility to determine the
masses of both the sleptons and the neutralino, as well as the slepton spin.
However, it will be shown that by using modified cuts and measurements at different
center-of-mass energies, these quantities can still be obtained with high
precision, as long as the slepton-neutralino mass difference is large enough.


\section{Mass measurement}
\label{mass}

It is well known that the masses of the sleptons and neutralino in the processes
\eqref{smu} and \eqref{sel} can be obtained from the endpoints $E_{\rm min,max}$ 
of the energy
spectrum of the final-state leptons $\ell=e,\mu$,
\begin{equation}
E_{\rm min,max} = \frac{\sqrt{s}}{4}
 \biggl (1-\frac{m_{\neu_1}^2}{m_{\sll}^2} \biggr )
 \Bigl (1\pm \sqrt{1-4m_{\sll}^2/s}\Bigr ),
\qquad
\begin{aligned}[c]
m_{\sll}^2 &= s\,\frac{E_{\rm min}E_{\rm max}}{(E_{\rm min}+E_{\rm max})^2},
\\
m_{\neu_1}^2 &= m_{\sll}^2 \biggl (1-
 \frac{2(E_{\rm min}+E_{\rm max})}{\sqrt{s}}  \biggr ).
\end{aligned}
\label{eminmax}
\end{equation}
To clearly observe both the upper and lower endpoints, the SM background must be
reduced to a sufficiently low level. This has been studied in detail for the
International Linear Collider (ILC), which is a proposed $e^+e^-$ collider with
up to 1~TeV center-of-mass energy \cite{ilccuts,cdr}\footnote{Backgrounds from
supersymmetry itself will not be considered here, but it has been shown that
they can be controlled with a few additional cuts \cite{cdr,susybkgd}.}. 
A typical set of cuts,
rescaled for a 3~TeV collider, is given by
\begin{align}
&\text{Exactly one $\ell^+$ and one $\ell^-$ with }
 (|\cos\theta_{\ell}| < 0.95\phantom{998} \text{ and } E_{\ell} > \text{25 GeV}), \label{ncut} \\[1ex]
&\begin{array}{@{}rl}
\text{No other visible object with either} & (|\cos\theta| < 0.95\phantom{998} \text{ and } E > \text{25 GeV}),
\\
\text{or} & (|\cos\theta| < 0.99998 \text{ and } E > \text{200 GeV}),
\end{array} \label{ecut} \\[1ex] 
&(p_{\ell^+} + p_{\ell^-})^2 > (\text{150 GeV})^2, \qquad 
p_{\rm miss}^2 > (\text{150 GeV})^2, \label{mcut} \\[1ex] 
&\cos\theta_{\ell^-} < +0.6, \qquad \cos\theta_{\ell^+} > -0.6, \label{acut} \\[1ex] 
&|\cos (\phi_{\ell^+} - \phi_{\ell^-})| < 0.9, \label{xcut}
\end{align}
where $E_{\ell^\pm}$, $p_{\ell^\pm}$, $\theta_{\ell^\pm},$ and $\phi_{\ell^\pm}$
are the energy, momentum, polar angle, and azimuthal angle of the (anti-)lepton,
respectively. The cut \eqref{ncut} ensures that two sufficiently energetic
oppositely charged leptons  are observed in the central detector. For the
purpose of the present study, the momenta of photons within a cone of $\Delta R
= \sqrt{(\Delta \eta)^2+ (\Delta \phi)^2} < 0.2$ around a lepton are added to
that lepton's momentum.  A veto \eqref{ecut} is imposed on any other objects
either in the central detector or in the low-angle calorimeter (LCAL), which for
a typical ILC detector design go down to a polar angle of about 5~mrad
\cite{ild}\footnote{In eq.~\eqref{ecut} a slightly larger cutoff of 6~mrad is
used to make sure that objects emitted with this  angle are fully contained in
the LCAL.}. The coverage of small polar angles is crucial for rejecting the
two-photon background \cite{martyn04}. The requirement \eqref{mcut} reduces
background from $Z$-boson production, while \eqref{acut} suppresses $W$-boson
background, which predominantly leads to leptons/anti-leptons in the
forward/backward region, respectively. Finally, the cut \eqref{xcut} rejects
back-to-back event topologies, which mainly stem from $W$ and $\tau$ pair
production and two-photon collisions.

\begin{table}[tb]
\renewcommand{\arraystretch}{1.2}
\centering
\begin{tabular}{|l||r|r|r|}
\hline
detector setup  & ILC design & 6$^\circ$ shielding cone & 20$^\circ$ shielding cone\\
\hline\hline 
$\sm^+\sm^-$ & 3062 & 3114 & 2780 \\
\hline 
$\mu^+\mu^-$ & 0 & 89 & 167 \\
$\tau^+\tau^- \to \mu^+\mu^- + X$ & 0 & 0 & 0 \\
$\mu^+\mu^-\nu\nu$ & 731 & 731 & 684 \\
$\gamma\gamma \to \mu^+\mu^- + X$ & 153 & 6341 & 7416 \\
\hline\hline 
$\se^+\se^-$ & 460 & 471 & 450 \\
\hline 
$e^+e^-$ & 0 & 2 & 4 \\
$\tau^+\tau^- \to e^+e^- + X$ & 0 & 0 & 0 \\
$e^+e^-\nu\nu$ & 253 & 253 & 221 \\
$\gamma\gamma \to e^+e^- + X$ & 135 & 6095 & 7242 \\
\hline
\end{tabular}
\mycaption{Event numbers for slepton signal and SM backgrounds after selection
cuts \eqref{ncut}--\eqref{xcut}, for different
detector setups. The numbers correspond to the parameters $\sqrt{s} = {}$3~TeV,
$m_{\sll} = {}$1~TeV, $m_{\neu_1} = {}$0.6~TeV, and an integrated luminosity of
1000~fb$^{-1}$.}
\label{evn}
\end{table}

\begin{figure}[p]
\raisebox{5cm}{\makebox[0mm]{a)}}\hspace{1.5mm}%
\epsfig{figure=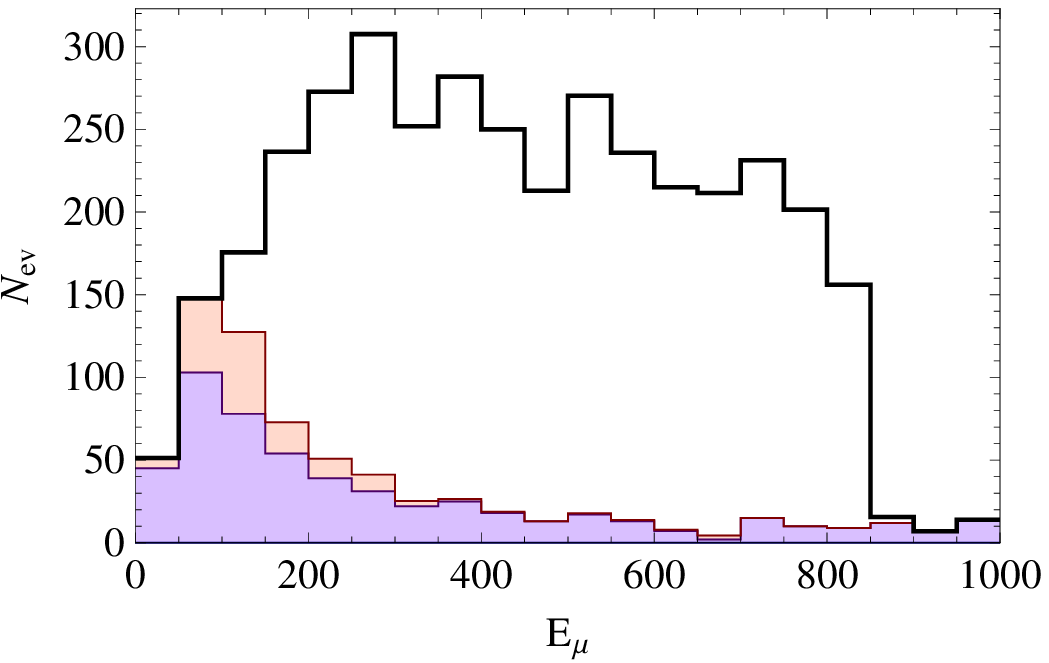, height=5cm}\rput[r](-0.6,4.5){\small ILC design}
\hfill
\raisebox{5cm}{\makebox[0mm]{b)}}%
\epsfig{figure=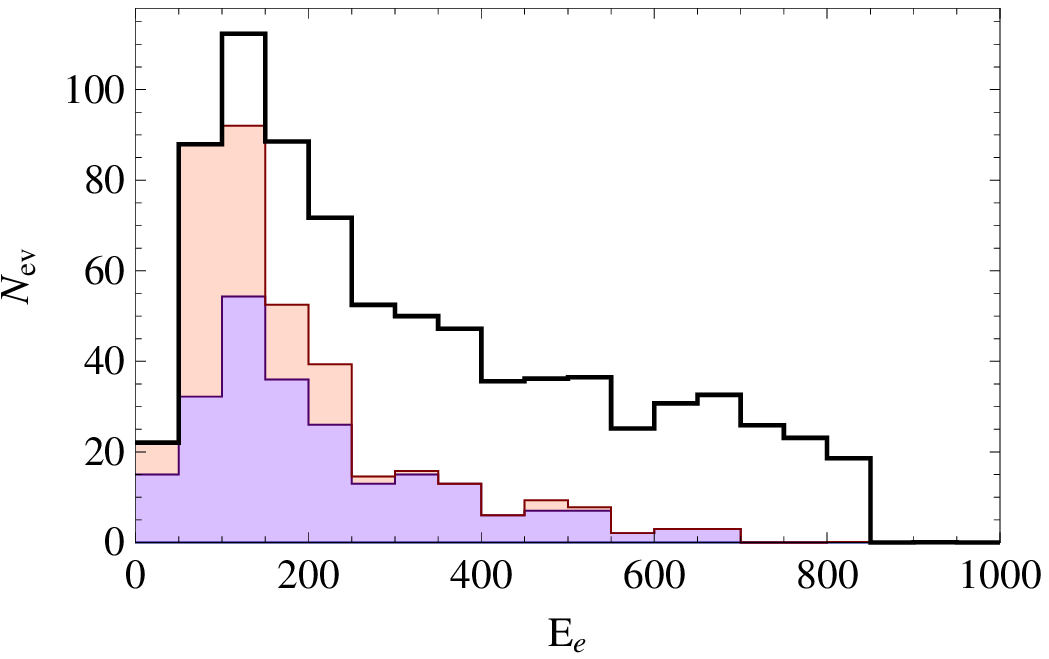, height=5cm}\rput[r](-0.6,4.5){\small ILC design}
\\[1ex]
\raisebox{5cm}{\makebox[0mm]{c)}}%
\epsfig{figure=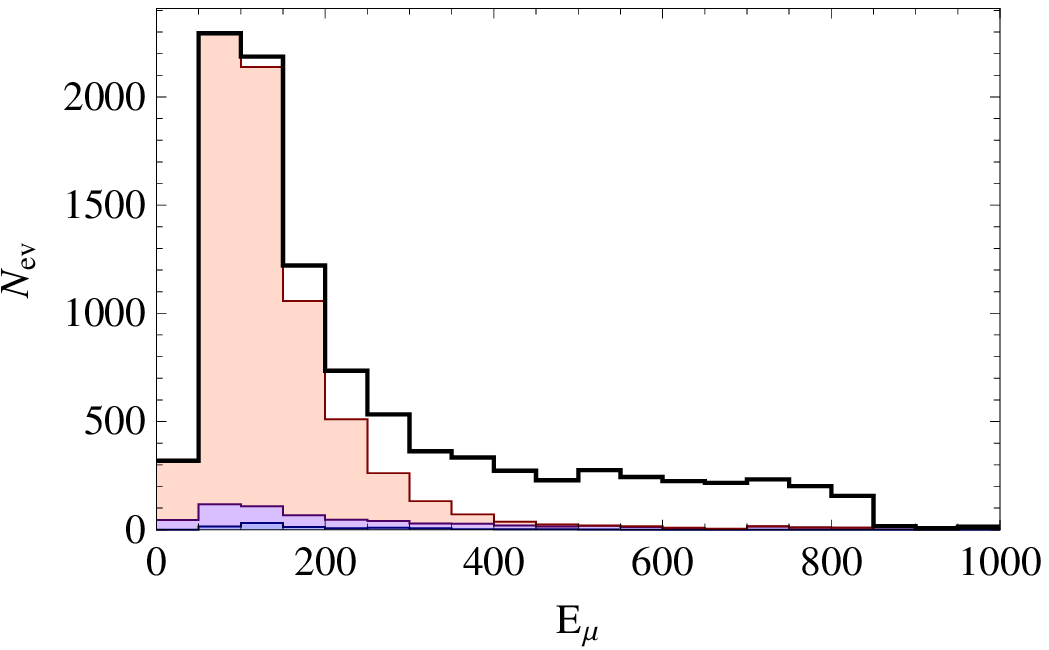, height=5cm}\rput[r](-0.6,4.5){\small 6$^\circ$ cone}
\hfill
\raisebox{5cm}{\makebox[0mm]{\hspace{4mm}d)}}%
\epsfig{figure=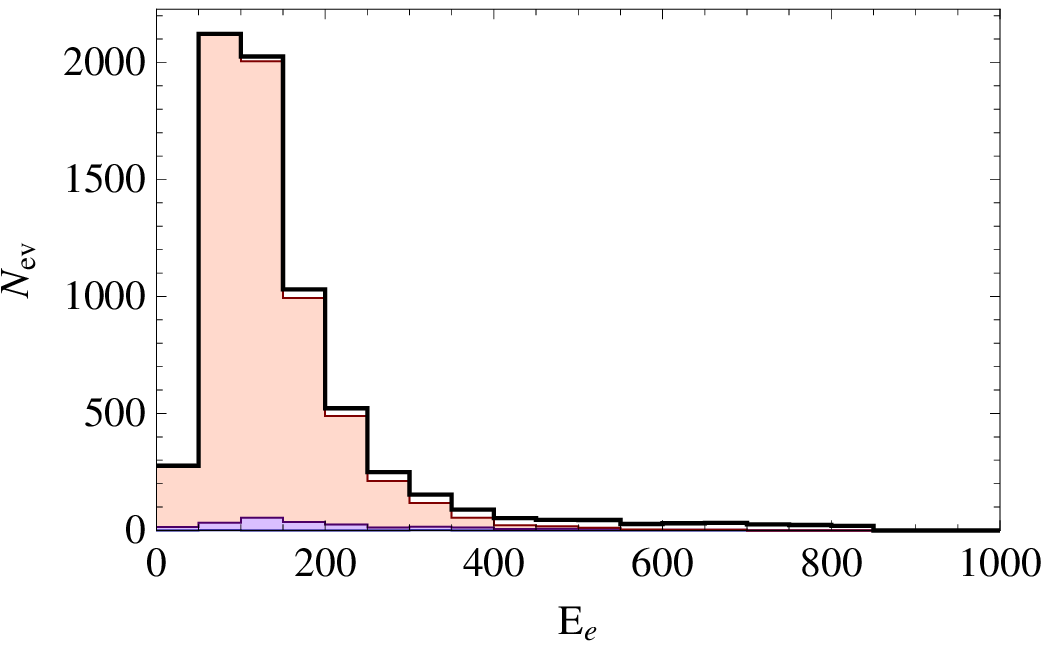, height=5cm}\rput[r](-0.6,4.5){\small 6$^\circ$ cone}
\\[1ex]
\raisebox{5cm}{\makebox[0mm]{e)}}%
\epsfig{figure=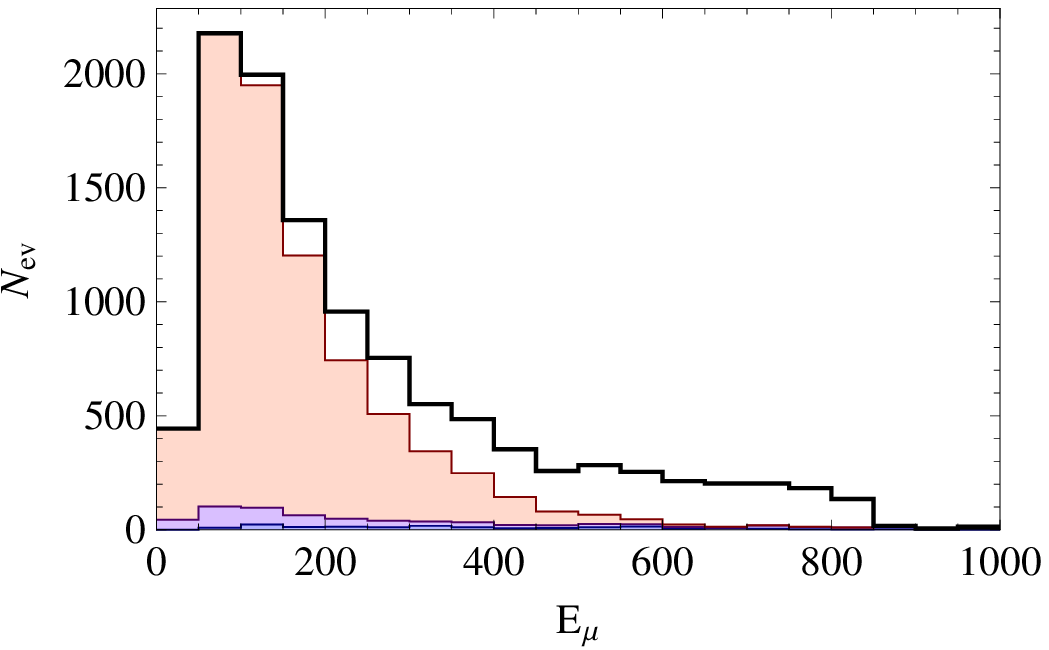, height=5cm}\rput[r](-0.6,4.5){\small 20$^\circ$ cone}
\hfill
\raisebox{5cm}{\makebox[0mm]{\hspace{4mm}f)}}%
\epsfig{figure=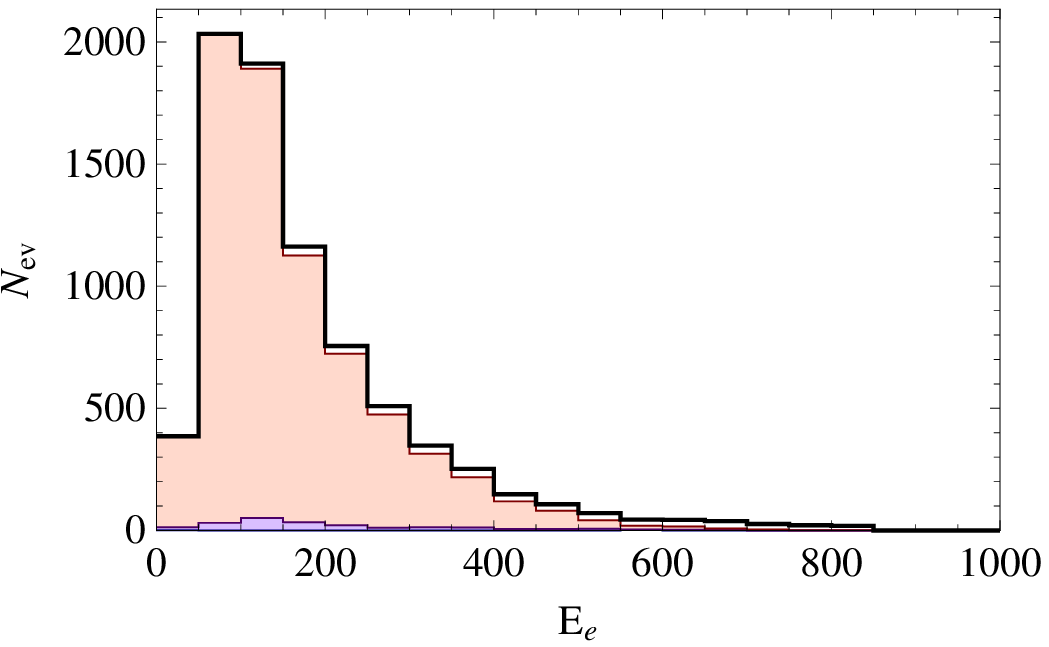, height=5cm}\rput[r](-0.6,4.5){\small 20$^\circ$ cone}
\\[2ex]
\newrgbcolor{darkblue}{0 0 0.5}
\newrgbcolor{lightblue}{0.7 0.7 1}
\newrgbcolor{darkmag}{0.3 0 0.4}
\newrgbcolor{lightmag}{0.85 0.75 1}
\newrgbcolor{darkred}{0.5 0 0}
\newrgbcolor{lightred}{1 0.85 0.8}
\anc\hspace{13mm}
\psframe[linecolor=darkblue,fillstyle=solid,fillcolor=lightblue](0.2,0)(1,0.4)
\rput[l](1.2,0.2){$\ell^+\ell^-$}
\psframe[linecolor=darkmag,fillstyle=solid,fillcolor=lightmag](4.2,0)(5,0.4)
\rput[l](5.2,0.2){$\ell^+\ell^-\nu\nu$}
\psframe[linecolor=darkred,fillstyle=solid,fillcolor=lightred](8.2,0)(9,0.4)
\rput[l](9.2,0.2){$\gamma\gamma$}
\psframe[linewidth=1pt](12.2,0)(13,0.4)
\rput[l](13.2,0.2){$\sll^+\sll^-$}
\mycaption{Energy distributions of the final-state leptons after application of
the selection cuts \eqref{ncut}--\eqref{xcut} described in the text, for
$\sm^+\sm^-$ production ($\ell=\mu$, left
column) and $\se^+\se^-$ pair production ($\ell=e$, right column). The three
rows correspond to different detector
setups. The plots are based on the parameters $\sqrt{s} = {}$3~TeV,
$m_{\sll} = {}$1~TeV, $m_{\neu_1} = {}$0.6~TeV, and an integrated luminosity of
1000~fb$^{-1}$.}
\label{edist}
\end{figure}

Results for signal and background event numbers and lepton energy distributions
after application of the cuts \eqref{ncut}--\eqref{xcut} are shown in
Tab.~\ref{evn} and Fig.~\ref{edist}~(a,b), for $\sqrt{s} = 3$~TeV, $m_{\sll} = 1$~TeV,
and $m_{\neu_1} = 0.6$~TeV. The signal and backgrounds from direct di-lepton 
production and $\gamma\gamma$ collisions have been generated with
{\sc Pythia 6.4} \cite{pythia}, while {\sc CompHEP 4.5} \cite{comphep} was used
for  four-fermion backgrounds, including intermediate resonant single and pair
production of $W$ and $Z$ bosons.

Figure~\ref{edist}~(a) shows that for an ILC-like detector, both endpoints of the
muon energy spectrum from smuon pair production are clearly visible over the
background after cuts. The location of the endpoints agrees with the nominal
values $E_{\rm min} = 122$~GeV and $E_{\rm max} = 838$~GeV calculated from
eq.~\ref{eminmax}. For selectron pair production, due to the smaller cross
section, the signal-to-background ratio is lower, and the extraction of $E_{\rm
min}$ from the data would require a very careful analysis of the background.

\vspace{\medskipamount}

The situation changes dramatically for a typical MC detector with an
uninstrumented shielding cone around the beam pipe. Two scenarios are
considered: a more conservative design with a 20$^\circ$ shielding cone, and a
very aggressive design with a 6$^\circ$ cone. Due to the lack of angular
coverage, the veto cut \eqref{ecut} now only works for objects with scattering
angles larger than the cone size, so that a much bigger portion of the
two-photon background passes the selection cuts. As can be seen from
Tab.~\ref{evn} and Fig.~\ref{edist}~(c,d,e,f), the signal event rates are
completely swamped by the $\gamma\gamma$ background and the lower endpoint of
the $E_\ell$ spectrum is lost in the statistical noise, both for smuon and
selectron pair production.

\subsection{Analysis strategy using additional cuts}

\begin{figure}
\raisebox{5cm}{\makebox[0mm]{a)}}%
\epsfig{figure=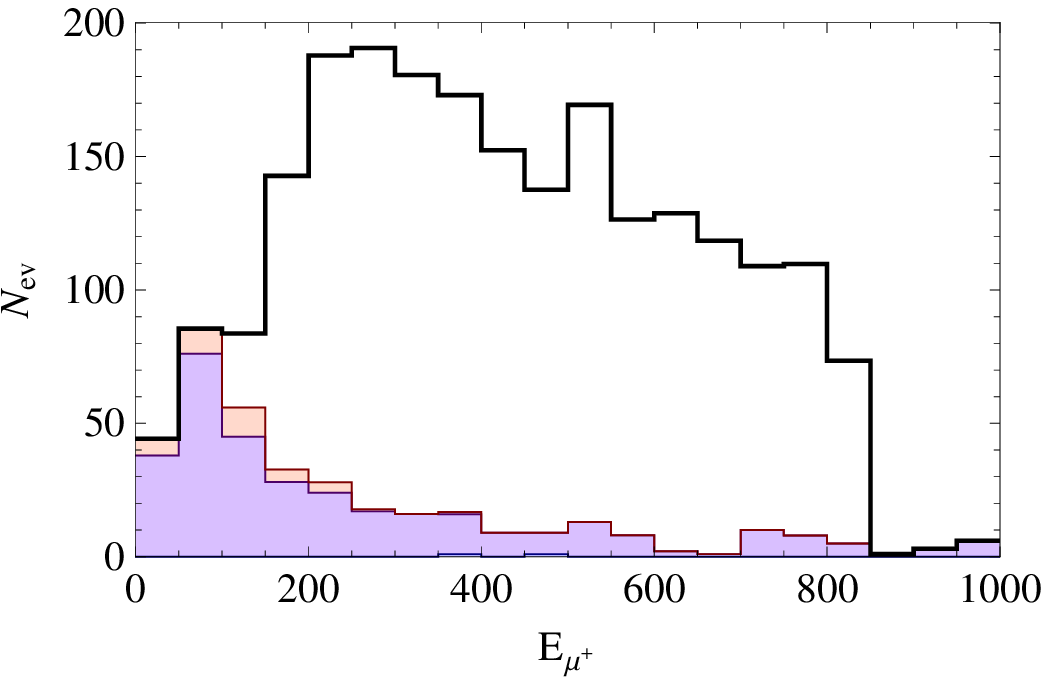, height=5cm}\rput[r](-0.6,4.5){\small 6$^\circ$ cone}
\hfill
\raisebox{5cm}{\makebox[0mm]{\hspace{4mm}b)}}%
\epsfig{figure=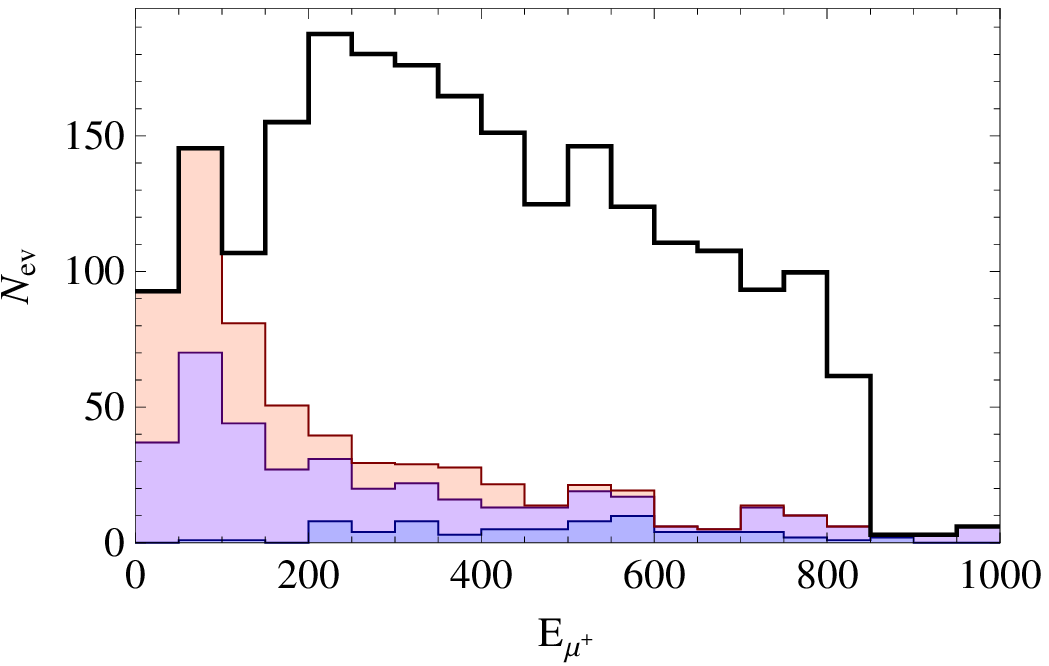, height=5cm}\rput[r](-0.6,4.5){\small 20$^\circ$ cone}
\\[1.2ex]
\newrgbcolor{darkblue}{0 0 0.5}
\newrgbcolor{lightblue}{0.7 0.7 1}
\newrgbcolor{darkmag}{0.3 0 0.4}
\newrgbcolor{lightmag}{0.85 0.75 1}
\newrgbcolor{darkred}{0.5 0 0}
\newrgbcolor{lightred}{1 0.85 0.8}
\anc\hspace{13mm}
\psframe[linecolor=darkblue,fillstyle=solid,fillcolor=lightblue](0.2,0)(1,0.4)
\rput[l](1.2,0.2){$\mu^+\mu^-$}
\psframe[linecolor=darkmag,fillstyle=solid,fillcolor=lightmag](4.2,0)(5,0.4)
\rput[l](5.2,0.2){$\mu^+\mu^-\nu\nu$}
\psframe[linecolor=darkred,fillstyle=solid,fillcolor=lightred](8.2,0)(9,0.4)
\rput[l](9.2,0.2){$\gamma\gamma$}
\psframe[linewidth=1pt](12.2,0)(13,0.4)
\rput[l](13.2,0.2){$\sm^+\sm^-$}
\mycaption{Energy distribution of the $\mu^+$ from $\sm^+\sm^-$ production and SM
backgrounds, after application of
the selection cuts \eqref{ncut}--\eqref{xcut} and \eqref{e2cut}. Results for a
6$^\circ$ (a) and a 20$^\circ$ (b) shielding cone are shown.
The plots correspond to the parameters $\sqrt{s} = {}$3~TeV,
$m_{\sm} = {}$1~TeV, $m_{\neu_1} = {}$0.6~TeV, and an integrated luminosity of
1000~fb$^{-1}$.}
\label{edist2}
\end{figure}

Figure~\ref{edist} also shows that the two-photon background falls off sharply with
increasing lepton energies and essentially disappears for $E_{e,\mu} \gesim
400$--500~GeV. This feature can be used to design a more effective cut for
removing the $\gamma\gamma$ background in a MC environment. By demanding that
one lepton, 
here arbitrarily chosen to be the negatively charged one,
has an energy above 400 GeV,
\begin{equation}
E_{\ell^-} > 400\gev, \label{e2cut}
\end{equation}
the signal-to-background ratio is much improved, see Tab.~\ref{evn2}. While this
cut obviously eliminates the lower part of the $E_{\ell^-}$ spectrum, the other
lepton, $\ell^+$, can still be used to measure the full lepton energy spectrum
with both the upper and lower endpoints $E_{\rm min,max}$, see
Fig.~\ref{edist2}.

\begin{table}[tb]
\renewcommand{\arraystretch}{1.2}
\centering
\begin{tabular}{|l||r|r|r|}
\hline
detector setup  & 6$^\circ$ shielding cone & 20$^\circ$ shielding cone\\
\hline\hline 
$\sm^+\sm^-$ & 1853 & 1616 \\
\hline 
$\mu^+\mu^-$ & 2 & 76 \\
$\tau^+\tau^- \to \mu^+\mu^- + X$ & 0 & 0 \\
$\mu^+\mu^-\nu\nu$ & 364 & 341 \\
$\gamma\gamma \to \mu^+\mu^- + X$ & 37 & 243 \\
\hline
\end{tabular}
\mycaption{Event numbers for smuon signal and SM backgrounds after selection
cuts \eqref{ncut}--\eqref{xcut} and \eqref{e2cut}, for different
detector setups. The numbers correspond to the parameters $\sqrt{s} = {}$3~TeV,
$m_{\sm} = {}$1~TeV, $m_{\neu_1} = {}$0.6~TeV, and an integrated luminosity of
1000~fb$^{-1}$.}
\label{evn2}
\end{table}

In an experimental analysis, the remaining two-photon background after cut
\eqref{e2cut} may be determined in a data-driven approach from the larger event
sample before application of \eqref{e2cut}, thus reducing the sizeable
theoretical uncertainties in the Monte-Carlo simulation of $\gamma\gamma$
collisions.
After subtraction of the residual SM background, the $\ell^+$ energy spectrum
can be fitted by a simple fit function with a sharp cutoff at $E_{\rm min}$ and
$E_{\rm max}$. For input parameters $\sqrt{s} = 3$~TeV,
$m_{\sll} = 1$~TeV, $m_{\neu_1} = 0.6$~TeV, and an integrated luminosity of
1000~fb$^{-1}$, the one-sigma errors from the fit are
\begin{align}
&\text{for 6$^\circ$ shielding cone:} &
&\delta m_{\sm,\rm fit} = {}^{+32}_{-40} \gev, &
&\delta m_{\neu_1,\rm fit} = {}^{+18}_{-14} \gev, \\
&\text{for 20$^\circ$ shielding cone:} &
&\delta m_{\sm,\rm fit} = {}^{+40}_{-46} \gev, &
&\delta m_{\neu_1,\rm fit} = {}^{+20}_{-18} \gev, 
\end{align}
where the numbers reflect statistical uncertainties from the signal and
subtracted background. Note that these errors are quite comparable to what one
would obtain with an ILC-like detector and \emph{without} the additional cut
\eqref{e2cut}:
\mbox{$\delta m_{\sm,\text{"ILC"}} = {}^{+31}_{-37} \gev$}, 
\mbox{$\delta m_{\neu_1,\text{"ILC"}} = {}^{+17}_{-14} \gev$}.
This seemingly counter-intuitive outcome
follows from the fact the mass errors are dominated by the fit of the lower
endpoint, whereas the cut \eqref{e2cut}, in combination with
\eqref{ncut}--\eqref{xcut}, primarily removes signal events with larger value of
$E_{\ell^+}$.

\subsection{Analysis strategy using two center-of-mass energies}

The results of the previous subsection look promising for the case of smuon pair
production. For selectrons, on the other hand, the pair production cross section
is only about 1/6 of the value for smuons. As a result, even with the cut
\eqref{e2cut}, the lower endpoint of the final-state electron energy spectrum is
still obscured by the statistical noise from the remaining SM background.

An alternative strategy, instead of trying to extract both the lower and upper
endpoints of the $E_\ell$ spectrum, is to use only the upper endpoint, but
at two different center-of-mass energies $\sqrt{s}_1$ and $\sqrt{s}_2$. From
these two measurements one can also uniquely determine the masses of the slepton
and of the neutralino, 
\begin{align} 
m_{\sll}^2 &= \frac{E_{\rm max,1} E_{\rm max,2} \,
 [(E_{\rm max,1}^2 + E_{\rm max,2}^2)\sqrt{s_1s_2} - 
 E_{\rm max,1} E_{\rm max,2} (s_1+s_2)]}{(E_{\rm max,1}^2 - E_{\rm max,2}^2)^2},
\label{s1s2a}
\\
m_{\neu_1}^2 &= m_{\sll}^2 \, \biggl [ 
 1+ 2\frac{E_{\rm max,1}^2 - E_{\rm max,2}^2}{E_{\rm max,1}\sqrt{s_1} -
   E_{\rm max,2}\sqrt{s_2}}
\biggr ],
\label{s1s2b}
\end{align}
where $E_{\rm max,1}$ and $E_{\rm max,2}$ are the upper endpoint energies
measured at $\sqrt{s_1}$ and $\sqrt{s_2}$, respectively.
This approach also allows one to drop the cut \eqref{e2cut}, which is only
helpful for improving the signal-to-background ratio at lower energies.

Using simulated events for $\sqrt{s_1}=2500$~GeV and $\sqrt{s_2}=3000$~GeV,
corresponding to
an integrated luminosity of 500~fb$^{-1}$ each, and fitting the upper endpoints
of the lepton energy distributions after cuts \eqref{ncut}--\eqref{xcut}, one
obtains from eqs.~\eqref{s1s2a}, \eqref{s1s2b}
\begin{align}
&\text{for 6$^\circ$ shielding cone:} &
&\delta m_{\sm,\rm fit} = {}^{+19}_{-16} \gev, &
&\delta m_{\neu_1,\rm fit} = {}^{+8}_{-6} \gev, \\
&\text{for 20$^\circ$ shielding cone:} &
&\delta m_{\sm,\rm fit} = {}^{+21}_{-18} \gev, &
&\delta m_{\neu_1,\rm fit} = {}^{+9}_{-7} \gev.
\end{align}
For selectron production, due to the smaller cross section, one obtains somewhat
larger errors:
\begin{align}
&\text{for 6$^\circ$ shielding cone:} &
&\delta m_{\se,\rm fit} = {}^{+49}_{-36} \gev, &
&\delta m_{\neu_1,\rm fit} = {}^{+21}_{-13} \gev, \\
&\text{for 20$^\circ$ shielding cone:} &
&\delta m_{\se,\rm fit} = {}^{+55}_{-32} \gev, &
&\delta m_{\neu_1,\rm fit} = {}^{+24}_{-13} \gev.
\end{align}
It is important, however, to note that this analysis strategy is very
robust, since the upper endpoint of the $E_\ell$ spectrum from
$\sll^+\sll^-$ production is essentially background-free and thus there are no
systematic uncertainties from the background simulation and subtraction.


\section{Reach}
\label{reach}

In the previous section, a scenario with a relatively large mass difference
$m_{\sm} - m_{\neu_1} = 400\gev$ was considered. For smaller mass differences,
which are expected in co-annihilation scenarios, the final-state lepton energy
spectrum becomes softer. For sufficiently small values of $m_{\sm} -
m_{\neu_1}$, the upper endpoint $E_{\rm max}$ can thus lie in a range with
significant SM background levels. In this case, both measurement strategies
described in the previous section will be affected by statistical and
systematical errors of the background, and a reliable measurement will be
impossible when these uncertainties become dominant.

\begin{figure}[t]
\centering\epsfig{figure=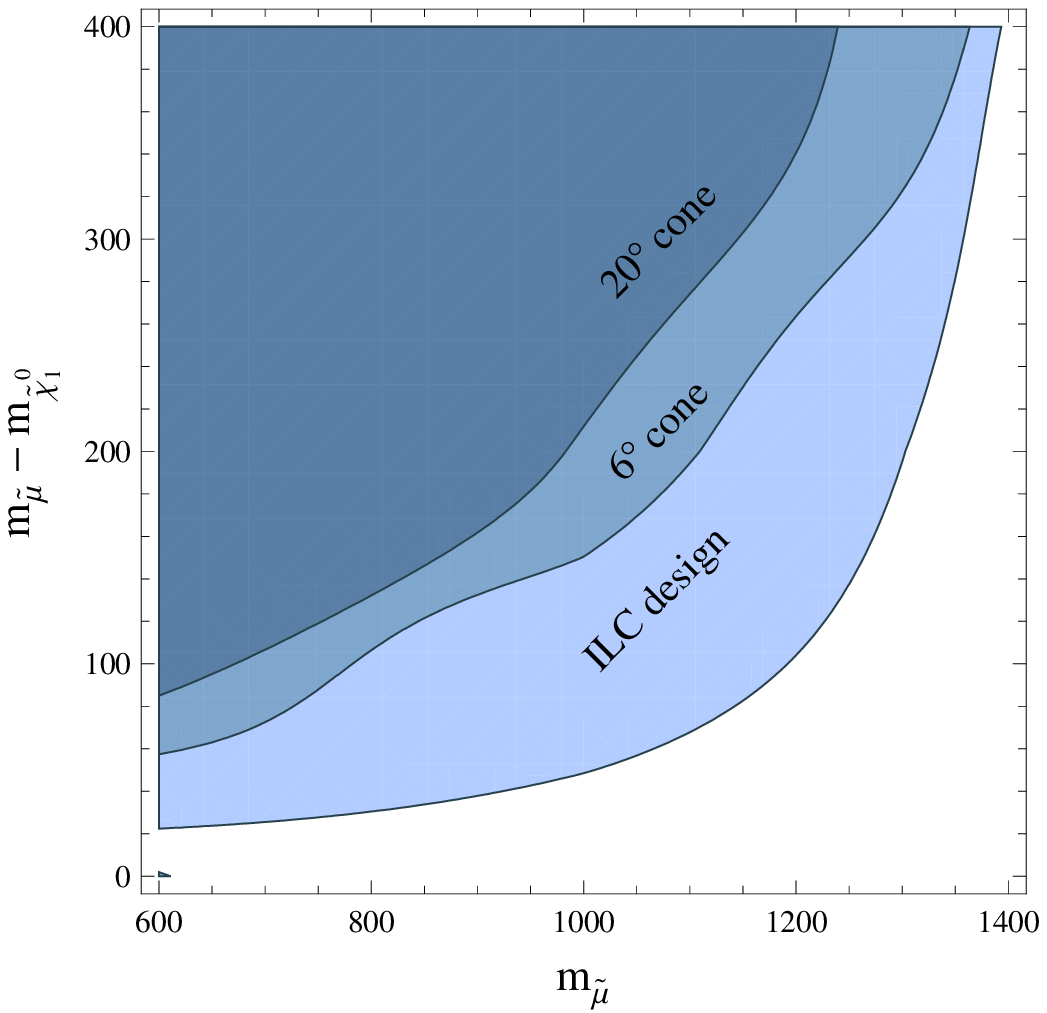, width=10cm}%
\vspace{-1ex}
\mycaption{Estimated region of smuon parameter space for which a reliable
measurement of the upper endpoint of the $E_\mu$ spectrum is possible,
for different detector setups. The criterion for what constitutes a reliable
measurement is described in the text. The plot corresponds to a muon collider with
$\sqrt{s} = 3$~TeV and an integrated luminosity of 1000~fb$^{-1}$.}
\label{reachp}
\end{figure}

To quantify this problem one needs to define a criterion for the minimum event
yield and allowable background level for the endpoint measurement. Here a very
simple rule is adopted, demanding that the bins of the energy spectrum near the
endpoint must have a signal-to-background ratio of at least one and must contain
at least 10 events per bin, for a bin width of 50~GeV (as in the figures above).
Using this criterion, the estimated reach of a 3~TeV muon collider is depicted
in Fig.~\ref{reachp}, for the three cases of an ILC-like detector, a 6$^\circ$
shielding cone, and a 20$^\circ$ shielding cone. The plot has been generated by
interpolating between simulated results for a range of smuon and neutralino
masses, applying the cuts \eqref{ncut}--\eqref{xcut} in each case.

As evident from the figure, the reach is limited by two factors. On one hand,
for smuon masses close to the beam energy the $\sm^+\sm^-$ cross section is
rather small and thus are affected by the high-energy tail of the residual SM
background. On the other hand, for small mass differences $m_{\sm}-m_{\neu_1}$,
the signal $E_\mu$ spectrum becomes soft and therefore lies in a kinematic
regime with large SM backgrounds. Since the background level grows with
increasing blind cone of the detector, the accessible parameter region is
reduced in both directions for a MC, compared to a hypothetical multi-TeV
ILC-like collider. In particular, it will be difficult to explore
co-annihilation scenarios at a MC.

Depending on the supersymmetry scenario, however, it may still be possible to
obtain a robust $\sm$ signal in this case if the smuons are also produced in the
decays of heavier superpartners. Such cascade decays can lead to a distinct
signal over the SM background as long as the visible decay products are
sufficiently energetic, $i.\,e.$ for the mass hierarchy $m_{\sm} \ll
m_{\tilde{X}} < \sqrt{s}$, where $m_{\tilde{X}}$ is the mass of the heavier
superpartner. Since this option is highly model-dependent, it will not be explored
further here.


\section{Spin determination}
\label{spin}

While the endpoints of the $E_{\ell}$ spectrum are related to the slepton and
neutralino masses, the shape of this distribution contains information about the
slepton spin. Since sleptons are scalars the $E_{\ell}$ distribution is
theoretically exactly flat, although in practice the shape is slightly distorted
by selection cuts.

\begin{figure}
\raisebox{5cm}{\makebox[0mm]{a)}}%
\epsfig{figure=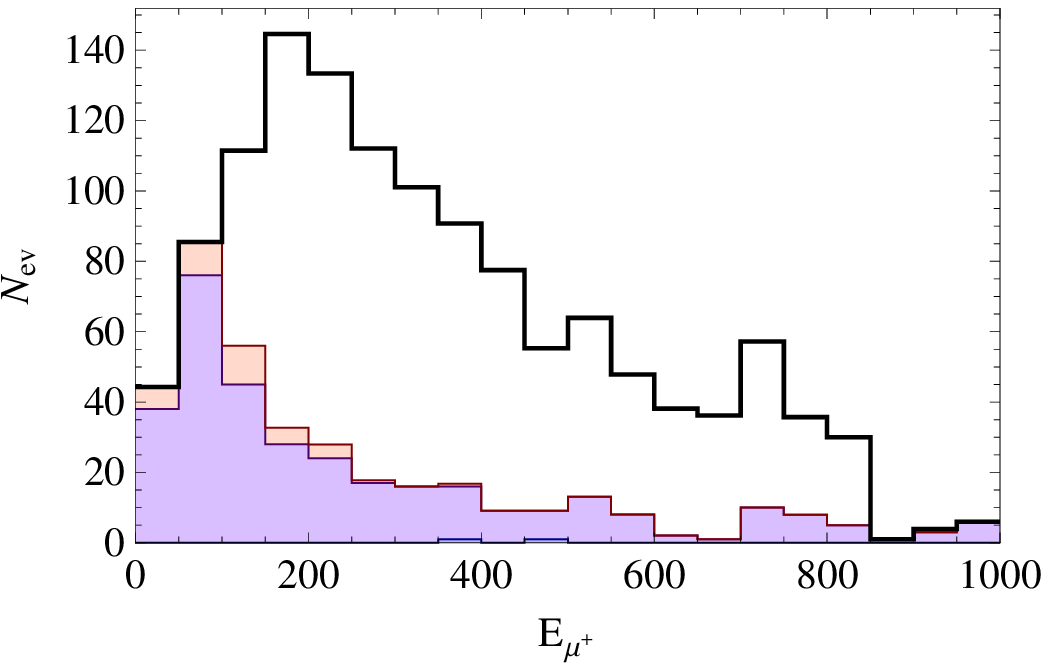, height=5cm}\rput[r](-0.6,4.5){\small 6$^\circ$ cone}
\hfill
\raisebox{5cm}{\makebox[0mm]{\hspace{4mm}b)}}%
\epsfig{figure=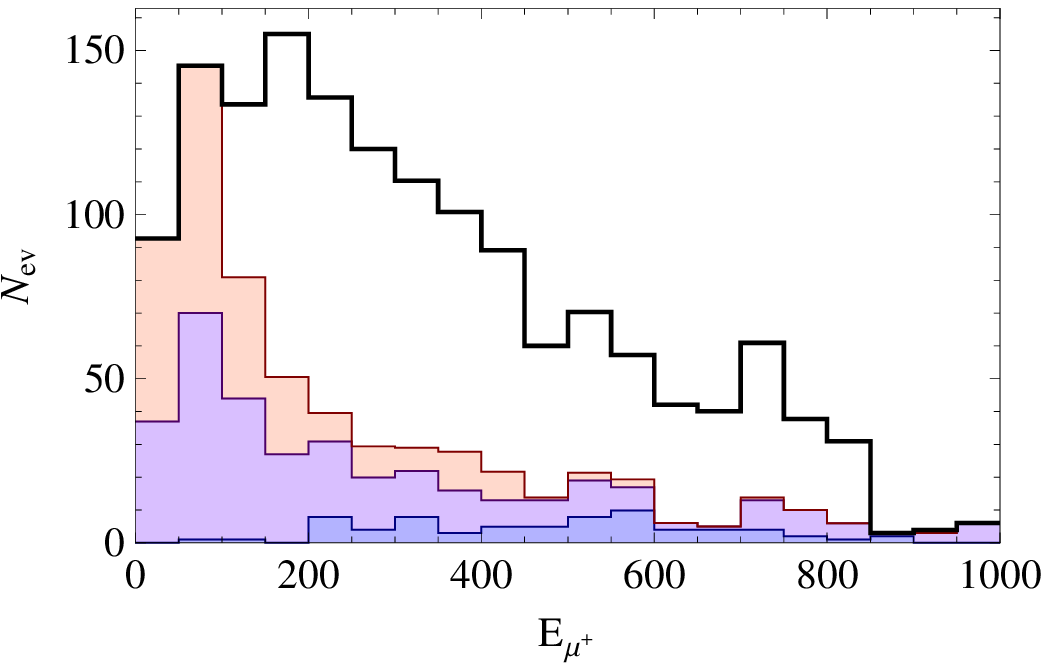, height=5cm}\rput[r](-0.6,4.5){\small 20$^\circ$ cone}
\\[1.2ex]
\newrgbcolor{darkblue}{0 0 0.5}
\newrgbcolor{lightblue}{0.7 0.7 1}
\newrgbcolor{darkmag}{0.3 0 0.4}
\newrgbcolor{lightmag}{0.85 0.75 1}
\newrgbcolor{darkred}{0.5 0 0}
\newrgbcolor{lightred}{1 0.85 0.8}
\anc\hspace{13mm}
\psframe[linecolor=darkblue,fillstyle=solid,fillcolor=lightblue](0.2,0)(1,0.4)
\rput[l](1.2,0.2){$\mu^+\mu^-$}
\psframe[linecolor=darkmag,fillstyle=solid,fillcolor=lightmag](4.2,0)(5,0.4)
\rput[l](5.2,0.2){$\mu^+\mu^-\nu\nu$}
\psframe[linecolor=darkred,fillstyle=solid,fillcolor=lightred](8.2,0)(9,0.4)
\rput[l](9.2,0.2){$\gamma\gamma$}
\psframe[linewidth=1pt](12.2,0)(13,0.4)
\rput[l](13.2,0.2){$\cha^+_1\cha^-_1$}
\mycaption{Muon energy distribution from $\cha^+_1\cha^-_1$ production and SM
backgrounds, after application of
the selection cuts \eqref{ncut}--\eqref{xcut} and \eqref{e2cut}. 
Results for a 6$^\circ$ (a) and a 20$^\circ$ (b) shielding cone are shown.
The plots correspond to the parameters $\sqrt{s} = {}$3~TeV,
$m_{\cha^\pm_1} = {}$1~TeV, $m_{\tilde{\nu}_\mu} = {}$0.6~TeV, and an integrated luminosity of
1000~fb$^{-1}$.}
\label{chap}
\end{figure}

As an alternative, let us consider the process
\begin{equation}
\mu^+\mu^- \to \cha^+_1\cha^-_1 \to \ell^+ \tilde{\nu}_{\ell} \ell^-
 \tilde{\nu}_{\ell}^*,
\end{equation}
where it is assumed that the sneutrinos $\tilde{\nu}_{\ell}^{(*)}$ decay
invisibly \cite{chasneu}. In this case, the primary pair-produced particles (the
charginos $\cha^\pm_1$) are fermions. For this process, the $E_{\ell}$
distribution is not flat, but spin correlations lead
to a relative accumulation of events near the lower end,
see Fig.~\ref{chap}.

In general, the $\cha^+_1\cha^-_1$ production cross section is different from
the $\sm^+\sm^-$ and $\se^+\se^-$ cross section, and this fact may also be
employed to distinguish these processes. However, the observable production rate
also depends on other model-dependent factors such as branching fractions.
In this work, therefore, only the shape information of the $E_{\ell}$
distribution will be used for spin discrimination, and the signal distribution
in Fig.~\ref{chap} has been rescaled to match the generator-level cross section
for $\sm^+\sm^-$ production.

Since the distinctive feature between scalars and fermion is the absence or
presence of an accumulation near the lower end of the $E_{\ell}$ distribution,
it is important to observe the whole $E_{\ell}$ spectrum, not only the upper
endpoint. To sufficiently reduce the SM backgrounds, one thus has to use the
selection cut \eqref{e2cut} in addition to \eqref{ncut}--\eqref{xcut}. Results
for $\ell=\mu$ are shown in Fig.~\ref{edist2} for scalars and Fig.~\ref{chap}
for fermions.

The discriminative power between the two spin scenarios can be illustrated with
a simple binned $\chi^2$ test, applied to the range 200--1000~GeV of the $E_\mu$
distribution. By demanding a lower limit of 200~GeV, the $\gamma\gamma$
background becomes subdominant, and the remaining SM background components can
be computed reliably from Monte-Carlo simulations. In the fit, the overall
normalization has been left free-floating. Assuming a total luminosity of
1000~fb$^{-1}$, the scalar and fermion scenarios can be distinguished with a
statistical significance of 7.7 standard deviations for a 6$^\circ$ blind cone
and 6.0 standard deviations for a 20$^\circ$ blind cone.

Note, however, that the spin determination is much more difficult or even
practically impossible for smaller mass differences $\Delta
m=(m_{\sm}-m_{\neu_1}),\,(m_{\cha^\pm_1}-m_{\tilde{\nu}})$, since in this case
the separation of signal and SM backgrounds becomes problematic, as
discussed in section~\ref{reach}. In particular, the cut \eqref{e2cut} will
entirely  remove all signal events for small values of $\Delta m$.


\section{Conclusions}
\label{concl}

At a future muon collider, the analysis of new-physics signatures with missing
energy is substantially impacted by the presence of uninstrumented shielding
cones around the beam pipe in the detector. Most notably, the limited angular
coverage of the detector leads to large backgrounds from two-photon collisions.
In this article, this problem has been studied in detail by considering the pair
production of sleptons, supersymmetric partners of leptons, as an example.
It was found that the potential for accessing this signal depends essentially on
the mass difference $\Delta m$ between the sleptons and their decay products,
which here are assumed to be stable neutralinos.

For relatively large mass differences, $\Delta m \gesim {\cal O}$(100~GeV), the
two-photon and other Standard Model backgrounds can be reduced by suitable
selection cuts, which allows one to clearly identify the new-physics signal and
determine properties such a slepton spin. Furthermore, by taking data at
two different center-of-mass energies, the masses of both the sleptons and their
decay products can be measured with percent-level precision.

On the other hand, for small slepton-neutralino mass differences, it becomes
difficult to robustly separate the photon-photon background from the signal. The
large background levels can make it impossible to determine the slepton masses
and spins or even establish a discovery of the signal in this case. In
consequence, the reach for new-physics signatures with small $\Delta m$ is more
limited at a muon collider compared to an $e^+e^-$ collider. However, it would
still be possible to analyze slepton production and decay in such a scenario at
a muon collider if one can take advantage of cascade decays of heavier
superpartners into the sleptons.

Interestingly, it turns out that the size of the blind shielding cones has only
a mild influence on these results, even when comparing two extreme scenarios: 
a very optimistic case with a $6^\circ$ shielding cone, and a
conservative case with a $20^\circ$ shielding cone.


\section*{Acknowledgements}

This project was supported in part by the National Science Foundation under
grant PHY-0854782.



\begin{thebibliography}{99}
\frenchspacing

\bibitem{mc}
For a recent summary, see $e.\,g.$
  V.~Shiltsev,
  Mod.\ Phys.\ Lett.\  A {\bf 25}, 567 (2010).

\bibitem{beamback}
  S.~Kahn, contribution to the {\it Low-Emittance Muon Collider Workshop,
  Fermilab, Batavia, IL (2006)}, {\tt
  http://www.muonsinc.com/mcwfeb06/\linebreak[0]%
  presentations/\linebreak[0]SKahn\_02072006\_DetectorBackgroundsLEMC.pdf};\\
  N.~V.~Mokhov, Y.~I.~Alexahin, V.~V.~Kashikhin, S.~I.~Striganov and A.~V.~Zlobin,
  contribution to the {\it 2011 Particle Accelerator Conference (PAC'11), New
  York, NY (2011)}, FERMILAB--CONF--11--094--APC;\\
  S.~A.~Kahn, M.~A.~C.~Cummings, T.~J.~Roberts, A.~O.~Morris, D.~Hedin and J.~Kozminski,
  contribution to the {\it 2011 Particle Accelerator Conference (PAC'11), New
  York, NY (2011)}, PAC--2011--THPO88.

\bibitem{klasen}
  J.~C.~Gallardo {\it et al.},
  ``Muon-Muon Collider: A Feasibility Study,'' in
{\it The Proceedings of 1996 DPF/DPB Summer Study on New Directions for
High-Energy Physics (Snowmass 96), Snowmass, CO}, chapt.~9;\\
  M.~Klasen,
  AIP Conf.\ Proc.\  {\bf 435}, 495 (1998).
  
\bibitem{mcued}
  D.~Greenwald and A.~Caldwell,
  contribution to the {\it 2011 Particle Accelerator Conference (PAC'11), New
  York, NY (2011)}, PAC--2011--MOP018.
  
\bibitem{ilccuts}
  K.~Hidaka, H.~Komatsu and P.~Ratcliffe,
  Nucl.\ Phys.\  B {\bf 304}, 417 (1988);\\
  T.~Tsukamoto, K.~Fujii, H.~Murayama, M.~Yamaguchi and Y.~Okada,
  Phys.\ Rev.\  D {\bf 51}, 3153 (1995);\\
  A.~Freitas, A.~von Manteuffel and P.~M.~Zerwas,
  Eur.\ Phys.\ J.\  C {\bf 34}, 487 (2004).

\bibitem{cdr}
  H.-U. Martyn, in {\it Conceptual Design of a 500 GeV $e^+e^-$ Linear Collider},
  eds. R.~Brinkmann, G.~Materlik, J.~Rossbach and A.~Wagner, DESY 1997--048 and
  ECFA 1997--182.

\bibitem{susybkgd}
  A.~Freitas, D.~J.~Miller and P.~M.~Zerwas,
  Eur.\ Phys.\ J.\  C {\bf 21}, 361 (2001).
     
\bibitem{ild}    
  T.~Abe {\it et al.}  [ILD Concept Group -- Linear Collider Collaboration],
  FERMILAB--LOI--2010--03, arXiv:1006.3396 [hep-ex].

\bibitem{martyn04}
  H.~U.~Martyn,
   in {\it Proc. of the International Conference on Linear Colliders (LCWS 04),
   Paris, France, 19-24 Apr 2004}
  [arXiv:hep-ph/0408226].
  
\bibitem{pythia}
  T.~Sj\"ostrand, S.~Mrenna and P.~Z.~Skands,
  JHEP {\bf 0605}, 026 (2006).

\bibitem{comphep}
  E.~Boos {\it et al.}  [CompHEP Collaboration],
  Nucl.\ Instrum.\ Meth.\ A {\bf 534}, 250 (2004).

\bibitem{chasneu}
  A.~Freitas, W.~Porod and P.~M.~Zerwas,
  Phys.\ Rev.\  D {\bf 72}, 115002 (2005).

\end{thebibliography}
\end{document}